\documentclass[a4paper,11pt]{amsart}

\DeclareRobustCommand{\SkipTocEntry}[5]{}

\usepackage{ amsmath }
\usepackage{marvosym}

\usepackage[T1]{fontenc}
\usepackage{empheq}
\usepackage{relsize}
\usepackage{bm}

\usepackage{upgreek}
\usepackage{ esint }
\usepackage{color}
\usepackage{comment}
\usepackage{amssymb}
\usepackage{amsfonts}
\usepackage{graphicx}

\usepackage{amscd}

\usepackage{slashed}
\usepackage{pdflscape}
\usepackage{tikz}
\usepackage{enumerate}
\usepackage{multirow}
\usepackage{stmaryrd}

\usepackage[linkcolor=blue]{hyperref}
\usepackage{mathrsfs}
\usepackage[colorinlistoftodos]{todonotes}

\usepackage[T1]{fontenc}

\definecolor{blue}{rgb}{.255,.41,.884} % RoyalBlue of svgnames
\definecolor{red}{rgb}{1, 0, 0} % Red of svgnames
\definecolor{green}{rgb}{.196,.804,.196} % LimeGreen of svgnames
\definecolor{yellow}{rgb}{1,.648,0} % Orange of svgnames
\definecolor{pink}{rgb}{1,0.5,0.5}

\newtheorem{theorem}{Theorem}[section]

\theoremstyle{definition}
\newtheorem{definition}[theorem]{Definition}
\newtheorem{example}[theorem]{Example}

\theoremstyle{remark}
\newtheorem{remark}[theorem]{Remark}

\usepackage{pdfsync}

\usepackage{graphicx} 
\usepackage{amssymb}

\newcommand{\be}{\begin{equation}}

\newcommand{\ee}{\end{equation}}

\newcommand{\ba}{\begin{array}}

\newcommand{\ea}{\end{array}}

\newcommand{\beq}{\begin{eqnarray}}

\newcommand{\eeq}{\end{eqnarray}}

\newtheorem{lm}{lemma}

\newtheorem{thee}{theorem}

\newtheorem{proo}{proposition}

\newtheorem{co}{corollary}

\newtheorem{rem}{remark}

\newtheorem{deff}{definition}

\newcommand{\bd}{\begin{deff}}

\newcommand{\ed}{\end{deff}}

\newcommand{\bl}{\begin{lm}}

\newcommand{\el}{\end{lm}}

\newcommand{\bp}{\begin{proo}}

\newcommand{\ep}{\end{proo}}

\newcommand{\bt}{\begin{thee}}

\newcommand{\et}{\end{thee}}

\newcommand{\bc}{\begin{co}}

\newcommand{\ec}{\end{co}}

\newcommand{\brm}{\begin{rem}}

\newcommand{\erm}{\end{rem}}

\usepackage{t1enc}

\newcommand{\newc}{\newcommand}

\let\ccdot.

\newc{\aR}{\mbox{\boldmath{$ R$}}}
\newc{\aS}{\mbox{\boldmath{$ S$}}}
\newc{\aT}{\mbox{\boldmath{$ T$}}}
\newc{\aW}{\mbox{\boldmath{$ W$}}}

\newc{\aD}{\mbox{\boldmath{$ D$}}\hspace{-.2mm}}

\newc{\aK}{\mbox{\boldmath{$ K$}}}
\newc{\aL}{\mbox{\boldmath{$ L$}}}
\usepackage{amssymb}
\usepackage{amscd}

\newc{\obstrn}[2]{B^{#1}_{#2}}

\newcommand{\rpl}                         % +) or <+
{\mbox{$
\begin{picture}(12.7,8)(-.5,-1)
\put(0,0.2){$+$}
\put(4.2,2.8){\oval(8,8)[r]}
\end{picture}$}}

\newcommand{\lpl}                         % (+ or +>
{\mbox{$
\begin{picture}(12.7,8)(-.5,-1)
\put(2,0.2){$+$}
\put(6.2,2.8){\oval(8,8)[l]}
\end{picture}$}}

\usepackage{ifthen}

\newc{\tensor}[1]{#1}
\newc{\Mvariable}[1]{\mbox{#1}}
\newc{\down}[1]{{}_{#1}}
\newc{\up}[1]{{}^{#1}}

\newc{\JulyStrut}{\rule{0mm}{6mm}}
\newc{\midtenPan}{\mbox{\sf S}}
\newc{\midten}{\mbox{\sf T}}
\newc{\midtenEi}{\mbox{\sf U}}
\newc{\ATen}{\mbox{\sf E}}
\newc{\BTen}{\mbox{\sf F}}
\newc{\CTen}{\mbox{\sf G}}

\def\sideremark#1{\ifvmode\leavevmode\fi\vadjust{\vbox to0pt{\vss
 \hbox to 0pt{\hskip\hsize\hskip1em
 \vbox{\hsize2cm\tiny\raggedright\pretolerance10000
  \noindent #1\hfill}\hss}\vbox to8pt{\vfil}\vss}}}

\numberwithin{equation}{section}

\newcommand{\hh}{{\hspace{.3mm}}}

\renewcommand\geq{\geqslant}

 \newcommand{\bdot }{{\mathop{\lower0.33ex\hbox{\LARGE$\cdot$}}}}

\newcommand{\superimpose}[2]{%
  {\ooalign{$#1\@firstoftwo#2$\cr\hfil$#1\@secondoftwo#2$\hfil\cr}}}

\begin{document}

\title{
{Contact Geometry and Quantum Mechanics
}}
\author{ G.~Herczeg${}^\sharp$ \&  Andrew Waldron${}^\natural$}

\address{${}^\sharp$
Department of  Physics,
  University of California,
  Davis, CA 95616, USA,}\email{\tt herczeg@ucdavis.edu}.
   
  \address{${}^{\natural}$
  Center for Quantum Mathematics and Physics (QMAP)\\
  Department of Mathematics\\ 
  University of California\\
  Davis, CA95616, USA} \email{wally@math.ucdavis.edu}

\vspace{10pt}

\renewcommand{\arraystretch}{1}

%Title of paper
\title{Contact Geometry and Quantum Mechanics}

%\author{G.~Herczeg}
%
%\affiliation{Department of  Physics,
%  University of California,
%  Davis, CA 95616, USA, {\tt herczeg@ucdavis.edu}.}
%
%
%
%
%\author{A.~Waldron}
%
%\affiliation{Department of Mathematics and the Center for Quantum Mathematics and Physics (QMAP), University of California, Davis, CA 95616, USA, {\tt wally@math.ucdavis.edu}.}

\date{\today}

\begin{abstract}
\noindent
We present a generally covariant  approach to quantum mechanics in which
generalized positions, momenta and {\it time} variables are treated as coordinates on a fundamental  ``phase-spacetime''.
%,  and
%dynamics are  given
%by a  phase-spacetime  contact  structure.
We show that 
this covariant
starting point 
makes quantization into a  purely geometric flatness condition. 
% The Schr\"odinger 
%equation is a parallel transport equation.
% Classical laboratories evolve along  the underlying phase-spacetime whose geometry
% allows comparison of the quantum system at different phase-spacetime points.
This  makes quantum mechanics purely geometric, and possibly even topological. Our approach
is especially useful for time-dependent problems 
and systems subject to ambiguities in choices of clock or observer.
As a byproduct, we give a
derivation 
and generalization of 
the Wigner functions of standard quantum mechanics.
\color{black}
 \end{abstract}

\maketitle

\tableofcontents

\section{Contact geometry} Mechanics is usually formulated in terms of an even $2n$-dimensional phase-space (or symplectic manifold) with time treated as an external parameter and dynamics determined by a choice of Hamiltonian. Yet classical physics ought not depend on choices of clocks.
However, 
Einstein's principle of general covariance 
can be applied to this situation by introducing an odd $(2n+1)$-dimensional phase-spacetime manifold~$Z$. Dynamics is now encoded by giving~$Z$ a (strict) contact structure---{\it i.e.},  a one-form $\alpha$ 
subject to a non-degeneracy condition
on the (phase-spacetime) volume form:
\begin{equation}\label{volume}
{\rm Vol}_\alpha :=\alpha\wedge (d\alpha)^{\wedge n}\neq 0\, .
\end{equation}
Physical phase-spacetime trajectories $\gamma$  are determined by extremizing the action 
\begin{equation}\label{action}
S=\int_\gamma \alpha\, .
\end{equation}
Since the integral of a one-form along a path $\gamma$ is a coordinate invariant quantity, general covariance (both worldline and target space) is built in from the beginning~\cite{Wald}. The equations of motion are
\begin{equation}\label{Reeb}
\varphi(\dot \gamma,\bdot)=0\, ,
\end{equation}
where the two-form 
$
\varphi:=d\alpha
$
is maximal rank by virtue of Eq.~\eqref{volume} and $\dot \gamma$ is a  tangent vector to the path $\gamma$ in $Z$. 

The structure $(Z,\alpha)$ is called a contact geometry and Eq.~\eqref{Reeb}
determines its Reeb dynamics~\cite{Geiges}. In addition to general covariance, this formulation of mechanics enjoys a Darboux theorem, which
implies the existence of {\it local} coordinates $(\psi,\pi_A,\chi^A)$ such that $\alpha=\pi_A d \chi^A-d\psi$ (where $A=1,\ldots,n)$ that trivialize the dynamics.
Hence one might hope to treat  classical {\it and} quantum mechanics as contact topology problems.

\medskip

\section{Goal}

We aim to develop a generally phase-spacetime covariant formulation of quantum mechanics. 
We find a formulation of quantum mechanics in terms of  intrinsic geometric structures on a contact 
manifold.
Our approach is 
similar to Fedosov's quantization of symplectic manifolds~\cite{Fed}, and indeed we were partly inspired by that work and subsequent BRST applications of Fedosov quantization\footnote{A contact analog of  Fedosov's connection for Poisson structures, where the fiber, rather than the base manfold, has a contact structure  was given in~\cite{Yoshioka}.} to models of higher spins~\cite{BRSTFed, Barn}. 
Quantization based on contact geometry has been studied before: For example, Rajeev~\cite{Rajeev} considers  quantization beginning with (classical) Lagrange brackets (the contact analog of Poisson brackets). Fitzpatrick~\cite{Fitz}  has extended this work to a rigorous geometric quantization setting.
There is also earlier work by Kashiwara~\cite{Kashi} 
that studies  sheaves of pseudodifferential operators over contact manifolds.
Investigations motivated by quantum cosmology  of the  so-called ``clock ambiguity'' in the quantum dynamics of time reparameterization invariant theories may be found in~\cite{Albrecht}. Contact geometry has also been employed in studies of choices of quantum clocks in~\cite{Warsaw}.

\medskip
\medskip

\section{BRST analysis} Because it is worldline diffeomorphism invariant, the system with action~\eqref{action} has one first class constraint. From the Darboux expression for the contact form~$\alpha$ we see that there are also $2n$ second class constraints (the canonical momenta for the coordinates~$\chi^A$ are constrained to equal the coordinates~$\pi_A$). The quantization of constrained systems is well understood, thanks to  the seminal work of Becchi, Rouet, Stora and Tyutin (BRST)~\cite{BRST}.
We employ the Hamiltonian BRST technology of Batalin, Fradkin and Vilkovisky (BFV)~\cite{BFV} as well as its 
extension to systems with second class constraints~\cite{BFVsecond}:

Let $z^i$ be phase-spacetime coordinates and introduce canonical momenta $p_i$ with Poisson brackets
$$
\{z^i,p_j\}_{\rm\tiny PB}=\delta_{\,j}^i\, .
$$
The second class constraints are
$$
C_i=p_i-\alpha_i\, ,
$$
where  $\alpha=\alpha_i dz^i$, $\varphi=\frac12 \varphi_{ij}dz^i\wedge dz^j$, and
$$
\{C_i,C_j\}_{\rm\tiny PB}=\varphi_{ij}\, .
$$
Second class constraints require 
 Dirac brackets; alternatively one may introduce $2n$ new variables $s^a$ with Poisson brackets
$$
\{s^a,s^b\}_{\rm \tiny PB}=J^{ab}\, ,
$$
where $J$ is a constant, maximal rank, $2n\times 2n$ matrix~\cite{BFVsecond}.  At least locally, we can introduce $2n$ linearly independent soldering forms~$e^a$ (analogous to the vielbeine/tetrads of general relativity)  such that
$$
\varphi =\tfrac{1}{2}J_{ab} e^a \wedge e^b\, ,
$$
and $J_{ab}J^{bc}=\delta_{\,a}^c$.
In these terms our system is now described by an extended action functional subject only to $2n+1$ first class constraints:
%\begin{widetext}
\begin{equation}\label{sext}
S_{\rm \tiny ext}[z(\tau),s(\tau)]=\int\Big[
\tfrac 12 s^a J_{ab} \dot s^b  +\dot z^i 
\big(\alpha_i(z) + s^a J_{ab} e_{i}{}^b(z)+ \omega_i(z,s)\big)\Big]d\tau\, .
\end{equation}
%\end{widetext}
In the above, $\tau$ is an arbitrary choice of worldline parameter, and the $s$-dependent one-form  
$\omega(z,s)$ on $Z$ must be chosen to obey~\footnote{Note that $\{\Omega\wedge\Omega\}_{\rm \tiny PB}:=dz^i \wedge dz^j\hh \{\Omega_i,\Omega_{j}\}_{\rm \tiny PB}$. 
In related work, the authors of~\cite{Albert} have constructed a
flat Cartan Maurer connection from a central extension of the group of canonical transformations.
}
$$
d\Omega+\frac12\, 
\{\Omega\wedge\Omega\}_{\rm \tiny PB}=0\, ,
$$ 
where $\Omega=\alpha + s^aJ_{ab} e^b + \omega$,
in order that the extended constraints $C_i^{\rm \tiny ext}=p_i-\Omega_i$ are first class. 
Locally, the Darboux theorem implies that a set of  one-forms $e^a$ with a flat connection exists.

The gauge invariances
$$
\delta z^i = \varepsilon^i(\tau)\, ,\qquad
\delta s^a= \varepsilon^i(\tau)J^{ab} \frac{\partial \Omega_i}{\partial s^b}\, ,
$$
ensure~\footnote{Here we assume that the rectangular matrix $\frac{\partial \Omega_i}{\partial s^b}$ has maximal rank, which is guaranteed at least in a  neighborhood of $s=0$.} that the equations of motion 
$$
J_{ab}\dot s^b +  \dot z^i \frac{\partial \Omega_i}{\partial s^a}= 0 = 
\dot z^i\big(\partial_i \Omega_j 
-\partial_j \Omega_i\big)
-
\dot s^a
\frac{\partial \Omega_j}{\partial s^a}\, ,
$$
are equivalent to Reeb dynamics.

Now that we are dealing with a first class constrained system, the BFV quantum action follows directly
$$
S_{\rm \tiny qu}=\int \big[\Theta + \{Q,\Phi\}_{\rm \tiny PB}\big]\, .
$$
Here $\Phi$ is the gauge fixing fermion for  some choice of gauge and $\Theta$ is the BRST-extended symplectic current
$$
\Theta = p_i \dot z^i  + \frac12 s^a J_{ab} \dot s^b
+ b_i \dot c^i\, ,
$$
where $(b_i,c^i)$ are canonically conjugate Grassmann ghosts. The BRST charge $Q=c^i C_i^{\rm \tiny ext}$ is determined by the first class constraints.

\medskip
\medskip
%\vspace{-.5cm}

\section{Quantization} We are now ready to quantize the contact formulation of classical mechanics.
The physical picture  underlying our method
  closely mimics general relativity:
Spinors in curved space are described by gluing a copy of a flat space Clifford algebra and its spin representation to each point in spacetime using vielbeine 
and the spin connection to compare spinors at differing spacetime points.
Mathematically, this is an example of a vector bundle in which context vielbeine are called soldering forms.
Here we want to glue a copy of standard quantum mechanics to each point $z$ in the phase-spacetime~$Z$,
which we view as the fibers of a suitable vector bundle, and then construct a connection $\nabla$ to compare differing fibers, as depicted below:

%\medskip
\medskip

%\vfill
%\eject

%\newpage

\vspace{0cm}
\begin{center}
\begin{picture}(100,100)(0,80)
\put(-20,55){
\includegraphics[width=6cm, height=4cm]
%scale=.3
{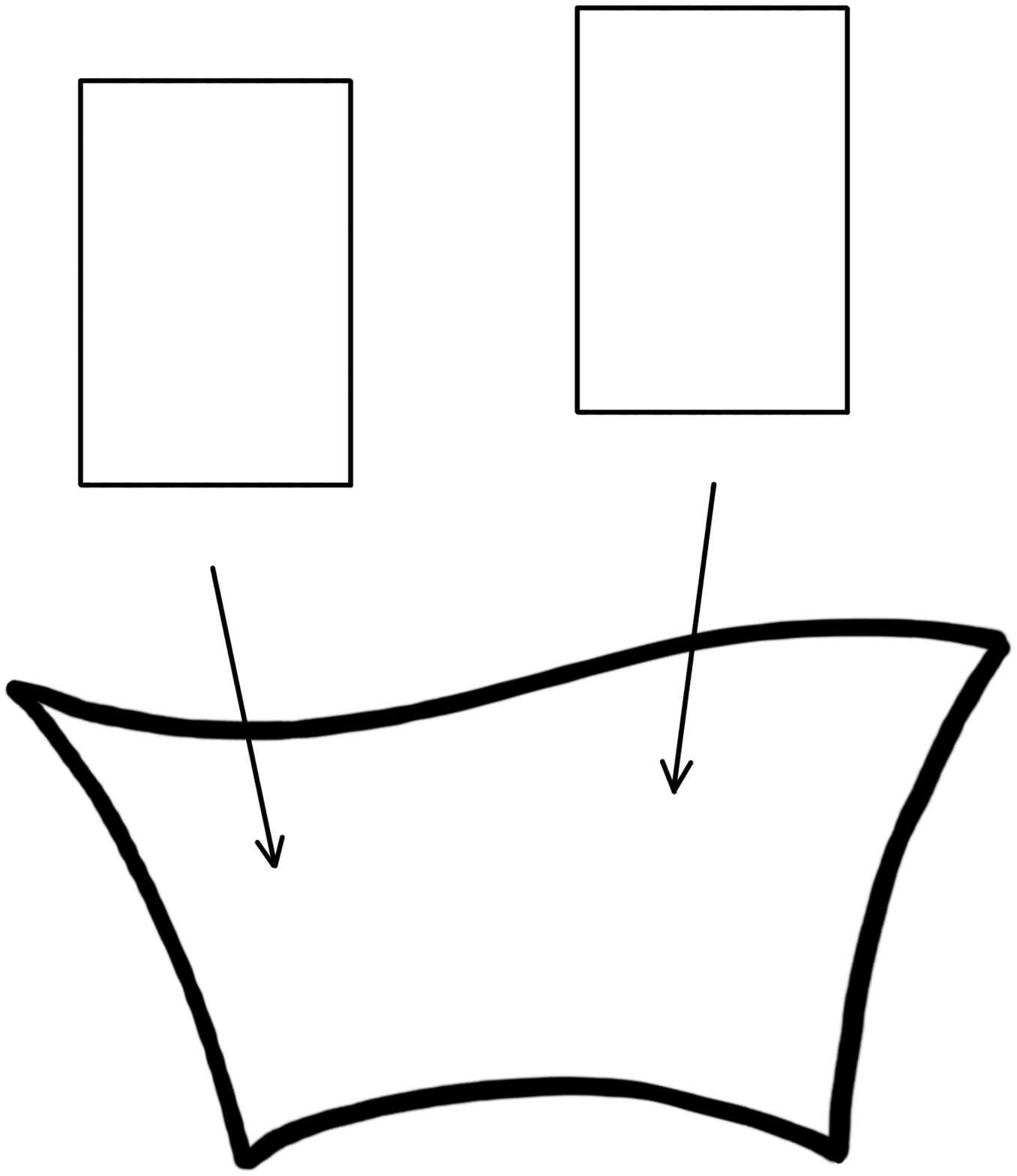}}
%{contact_art.jpg}}
%{contact_art.png}}
%\put(-50,5){\vector(1,0){200}}
\put(-50,167){\scalebox{.9}{Quantum Mechanics $F$}}
\put(90,47){\scalebox{.9}{Phase-spacetime $Z$}}
\put(40,82){$z$}
\put(80,85){$z'$}
\put(47,140){$\stackrel{\textstyle \nabla}{\longleftrightarrow}$}
\end{picture}
\end{center}
\vspace{1.8cm}

%\begin{center}
%INSERT OUR ARTWORK HERE
%%\includegraphics[width=9cm, height=6.5cm]{our_artwork.pdf}
%\end{center}

\medskip

\noindent
In this picture, quantum mechanics along the fibers is described in terms of the variables~$s^a$ which
are quantized in the standard way by choosing some polarization in which
$$
\hat s^a = \Big (S^A,\frac{\hbar}i\hh\frac\partial{\partial S^A}\Big)\, .
$$
Quantum wavefunctions $\Psi(S^A)$ depend on half the $s$-variables $S^A$ spanning  ${\mathbb R}^n$, and the inner product is the usual one: $\langle \Psi,\Psi'\rangle = \int_{\mathbb R^n} \Psi^*\Psi'$. 
The ``Schr\"odinger equation" along each fiber as well as the parallel transport of quantum mechanics from fiber to fiber is controlled by the connection $\nabla$ given by the quantum BRST charge~$\widehat Q^{\phantom{A^b}}\!\!\!\!\!\!\!$. To compute this connection, we quantize the contact coordinates~$z^i$ and their momenta using the polarization
$$
\hat p_i=\frac{\hbar}{i} \frac\partial{\partial z^i}\, .
$$
In addition, we identify the Grassmann ghosts $c^i$ with a basis of one-forms $dz^i$ along $Z$. Hence BRST wavefunctions  now depend on $(z^i, dz^i, S^A)$ and  may be viewed as differential forms on the contact manifold $Z$ taking values in the Hilbert space $L^2(\mathbb R^n)$.
The quantum BRST charge $ \widehat Q=\frac{\hbar}i\nabla$ where $\nabla$ is the operator-valued connection~\footnote{In fact, exactly such a connection over a symplectic manifold has been introduced in~\cite{Krysl}.} 
$$
\nabla = d - \frac i\hbar \hh \hh\widehat \Omega(z,\hat s)\, .
$$
Here the operator-valued one-form 
$$
\widehat \Omega=
\alpha + e_a \hat s^a + \widehat \omega(z,\hat s)\, ,
$$
where $\widehat\omega$ is an expansion in two and higher powers of~$\hat s$.
By nilpotence of $\widehat Q$, the operator~$\widehat \Omega$ is solved for by requiring that the connection $\nabla$ is flat:
$$
\nabla^2=0\, .
$$
Again, existence of a solution   on a local patch of $Z$ is guaranteed by the Darboux theorem.

Physical quantum states are given by the BRST cohomology at ghost number zero, {\it i.e.}, zero-forms $\Psi(z^i,S^A)$ subject to the parallel transport condition
$$
\nabla \Psi(z^i,S^A)=0\, .
$$
We then  search
for   solutions labeled by a set of quantum numbers ${\mathcal E}$
such that at each point $z\in Z$,
$\Psi_{\mathcal E}(z,S)$ are complete
and orthonormal:
\begin{equation}\label{inner}
\int \Psi^*_{\mathcal E'}(z,S)
\Psi_{\mathcal E}(z,S)\hh d^nS=\delta_{{\mathcal E},{\mathcal E'}}\, .
\end{equation}
Explicit solutions to this  condition for a broad class of 
models are given at the end of this letter. 

Finally we are ready to  build a set of generalized (contact covariant, non-diagonal) Wigner distributions encoding the quantum mechanical system: In a ket notation, we may think of the wavefunction $\Psi_{\mathcal E}(z,S)$ as a  state $|{\mathcal E};z\rangle$ labeled by quantum numbers~${\mathcal E}$ and indexed by control parameters $z$ given by points in the contact manifold $Z$.
In this notation, the display \eqref{inner} reads $\langle {\mathcal E}'\!\hh;z\hh |\hh {\mathcal E};z\rangle = \delta_{{\mathcal E},{\mathcal E'}}$. Then the generalized Wigner distribution is given by
$$
{\mathcal W}_{{\mathcal E},{\mathcal E'}}(z,z'):=
\langle {\mathcal E}'\!\hh;z'\hh\! |\hh {\mathcal E};z\rangle\, .
$$
The above expression can also be viewed as a two-point correlator for a field theory on the contact manifold. %\andrew{We can hit this with $\nabla$ or $\nabla'$...}

The Wigner distribution can be interpreted physically as follows. The control parameters $z^i$ correspond to 
values of  dials, knobs, meters {\emph{and clocks} in the classical laboratory, while ${\mathcal E}$ labels a quantum state prepared by the experimenter. 
Note that the parameters~$z$ are mutually commuting variables consistent with the phase space dependence of 
quantum mechanical Wigner functions.
The quantity
 $|W_{{\mathcal E},{\mathcal E'}}(z,z')|^2$ then measures the probability of observing the state ${\mathcal E}'$ given
 the state ${\mathcal E}$ was initially prepared and the classical laboratory
 has ``evolved'' in control parameter space from $z$ to $z'$.
 
\medskip
 
\section{Example:~Hamiltonian~mechanics}  Consider a
phase-spacetime with coordinates $(q^A,p_A,t)$ and
 contact form
$$
\alpha = p_A dq^A -H(p,q,t) dt\, .
$$
Choosing the worldline parameterization $t(\tau)=\tau$, the action principle of Eq.~\eqref{action} then gives Hamilton's equations for $(q^A,p_A)$ with Hamiltonian $H(p,q,t)$.
Here 
$$
\varphi = d\alpha = e_A\wedge f^A
$$
with
$$
e_A= dp_A+\frac{\partial H}{\partial q^A} \hh dt
\, ,\qquad
f^A= dq^A-\frac{\partial H}{\partial p_A}\hh dt
\, . 
$$
It is not difficult to formally solve order by order in the fiber operators
$\hat s^a=(S^A,\frac\hbar i\frac\partial{\partial S^A})$
for the operator-valued one-form~$\widehat \Omega$ and find a flat connection 
%(throughout the remainder of this letter we set $\hbar=1$)
%\begin{widetext}
$$
\nabla = d 
-\frac i\hbar \hh \alpha
+\frac i\hbar \hh  e_A S^A
 - f^A \frac{\partial}{\partial S^A} +\frac i\hbar \hh dt \sum_{\sigma\geq 2}\frac1{\sigma!}\frac{\partial^\sigma H(Z,t)}{\partial Z^{a_1}\cdots \partial Z^{a_\sigma}}
\hat s^{a_1} \cdots \hat s^{a_\sigma}\, ,
$$
%\end{widetext}
where $Z^a:=(q^A,p_A)$. 

%\newpage

The ``contact Schr\"odinger equation'' $\frac\hbar i \nabla \Psi=0$ now gives a triplet of equations 
$$
\left\{
\begin{array}{l}
\displaystyle
\frac\hbar i\Big[ \frac{\partial }{\partial q^A}-
\frac{\partial }{\partial S^A}\Big]\Psi
-p_A\Psi=0
\, ,\\[4mm]
\displaystyle
\frac\hbar i 
\frac{\partial \Psi}{\partial p_A}+S^A \Psi=0\, ,\\[4mm]
\displaystyle
\frac\hbar i
\frac{\partial \Psi}{\partial t}
+
i\sum_{\sigma\geq0}\frac1{\sigma!}\frac{\partial H}{\partial Z^{a_1}\cdots \partial Z^{a_\sigma}}\hat s^{a_1}\cdots \hat s^{a_\sigma}\Psi=0\, .
\end{array}
\right.
$$
The first two of these equations are solved via
$$
\Psi(q,p,t,s)=e^{-\frac{i}\hbar p_A S^A}\hh \psi(Q^A,t)\, ,
$$ 
where $Q^A=q^A+S^A$. The operator appearing in the summation in the third equation can be resummed to give $\widehat H=\big(H(Q,p+\frac\hbar i\frac \partial{\partial S})\big)_{\rm \tiny Weyl}$ where the symbols $Q^A$ and $\partial/\partial S^A$ are Weyl ordered. Thus $\psi(Q^A,t)$ obeys the time dependent Schr\"odinger equation
$$
i\hbar \hh\frac{\partial \psi(Q,t)}{\partial t} = \Big[H\Big(Q,\frac\hbar i\frac\partial{\partial Q}\Big)\Big]_{\rm \tiny Weyl}\hh  \psi(Q,t)\, .
$$
Focusing on the case where $\widehat H$ is time independent, the analysis is now standard: A complete orthonormal set of states $\{\psi_E(Q,t)=e^{-\frac i\hbar Et} \psi_E(Q)|E\in {\rm spec}(\widehat H)\}$ is labeled by energies $E$ (up to degeneracies). 
At equal values of $z=(q,p,t)$, the inner product of solutions
yields
$${\mathcal W}_{E,E'}(z,z)=
%\langle  \Psi_{E'}(z,S),\Psi_E(z,S)\rangle=
\delta_{E,E'}\, ,$$
in agreement with Eq.~\eqref{inner}, while at unequal values of the control parameters $z\neq z'$ but equal quantum numbers~$E=E'$,
$$
%\begin{multline*}
{\mathcal W}_{E,E}(z',z)=
e^{\frac i\hbar [E\delta t-\delta p_A \bar q^A]}
\int\frac{d^n\!S \hh d^n\!P}{(2\pi\hbar^2)^{\raisebox{.3mm}{$\scriptstyle n$}}}\,
e^{\frac i\hbar P_A \delta q^A}\,  W_E(S,P)\, ,
%\end{multline*}
$$
with $\bar q:=(q+q')/2$, $\delta p =p'-p$ and $\delta t=t'-t$. Here~$W_E$ is the Wigner function
$$
W_E(S,P)=\int d^ny\,  e^{\frac i\hbar P_A y^A} \psi_E^*(S-\tfrac y2)\hh\psi_E(S+\tfrac y2)\, .
$$
This is the fundamental building block of the phase space formulation of quantum mechanics~\cite{Phase space}.

\medskip

\section{Summary and discussion} 
By BRST quantizing  classical mechanics described  in terms of contact geometry,
we have reformulated  quantum mechanics 
as the parallel transport equation of a flat connection on a vector bundle over phase-spacetime. This implies that we have turned quantum mechanics into cohomology.
Our approach has a simple geometric  interpretation: (i) We maintain general covariance with respect to phase-space and time coordinates at all junctures, and (ii) we compare standard quantization at fixed phase-spacetime points using a connection, just as the Levi-Civita spin
 connection compares vectors, spinors {\it etc...}, from spacetime point to spacetime point. From a bundle viewpoint, this means that we quantize along fibers and compute inner products fiber-wise. Correlators are covariantly labeled by pairs of phase-spacetime points.
 This provides
 a derivation and generalization of  the Wigner functions 
 which are usually postulated in standard quantum mechanics.

Our approach is intrinsic to the data of a strict contact manifold, which is necessarily locally trivial.
This raises the tantalizing possibility that 
both classical and quantum dynamics 
can be completely described  
 in terms of the topological data of vector bundles over contact manifolds.

\medskip

%We will make the following correspondences
%
%\begin{center}
%\begin{tabular}{ccc}
%Quantum BRST charge $\widehat Q$ &
%$\longleftrightarrow$ & connection $\nabla$\\[1mm]
%Schr\"odinger equation
%&
%$\longleftrightarrow$ &
%parallel transport $\nabla \Psi = 0$\\[1mm]
%\end{tabular}
%\end{center}

%\begin{acknowledgments}
\noindent
\section*{Acknowledgments}
We thank James Conway, Mike Eastwood, Rod Gover, Jerry Kaminker, Emanuele Latini and Bruno Nachtergaele for discussions.
A.W. was supported in part by a 
Simons Foundation Collaboration Grant for Mathematicians ID 317562.
%\end{acknowledgments}

\appendix

\end{document}